\documentclass[showpacs,preprintnumbers,pre]{revtex4}
\usepackage{graphicx}
\usepackage{bm}

\begin{document}
\title{
Microstructure and velocity of field-driven Ising interfaces 
moving under a soft stochastic dynamic
}
\author{
Per Arne Rikvold$^{1,2,}$}\email{rikvold@csit.fsu.edu}
\author{M.~Kolesik$^{3,4,}$}\email{kolesik@acms.arizona.edu}
\affiliation{
$^1$Center for Materials Research and Technology, 
School of Computational Science and Information Technology, and 
Department of Physics,
Florida State University, Tallahassee, Florida 32306-4350\\
$^2$Center for Stochastic Processes in Science and Engineering, 
Department of Physics, Virginia Polytechnic Institute and State University,
Blacksburg, VA 24061-0435\\
$^3$Institute of Physics, Slovak Academy of Sciences,
Bratislava, Slovak Republic\\
$^4$Optical Sciences Center, University of Arizona,
Tucson, Arizona 85721
}
\date{\today}

\begin{abstract}
We present theoretical and dynamic Monte Carlo simulation results for the
mobility and microscopic structure of 1+1-dimensional Ising interfaces
moving far from equilibrium in an applied field under a single-spin-flip 
``soft'' stochastic dynamic. The soft
dynamic is characterized by the property that the effects of changes in
field energy and interaction energy factorize in the transition rate, in
contrast to the nonfactorizing nature of the traditional Glauber and
Metropolis rates (``hard'' dynamics). 
This work extends our previous studies of the Ising model with a hard
dynamic and the unrestricted SOS model with soft and hard dynamics. 
[P.~A.\ Rikvold and M.~Kolesik, J.\ Stat.\ Phys.\ {\bf 100}, 377 (2000); 
J.\ Phys.\ A {\bf 35}, L117 (2002); Phys.\ Rev.\ E {\bf 66}, 066116 (2002).] 
The Ising model with soft dynamics is found to have closely similar
properties to the SOS model with the same dynamic. In particular, the
local interface width does {\it not\/} diverge with increasing field, as
it does for hard dynamics. The skewness of the interface at nonzero
field is very weak and has the opposite sign of that obtained with hard
dynamics. 
\end{abstract}

\pacs{ 
05.10.ln 
68.35.Ct 
75.60.Jk 
68.43.Jk 
}

\maketitle


\section{Introduction}
\label{sec:Int}
The structure and dynamics of surfaces and interfaces significantly influence 
a host of material properties. Consequently, an 
enormous amount of work has been devoted to the study of  
moving and growing interfaces \cite{BARA95,MEAK98}. However, 
despite the fact that many important interface properties, such as mobility 
and chemical 
activity, are largely determined by the {\it microscopic\/} interface 
structure, the bulk of this effort has concentrated on large-scale scaling 
properties. 

Since the detailed physical mechanisms of the interface motion are most often 
unknown, it is useful to model the dynamic as a stochastic process 
defined by a set of transition probabilities. 
It is therefore important to gain better insight 
into how the driving force (such as an 
applied magnetic or electric field or a chemical-potential difference) may 
alter the microscopic interface structure for different stochastic dynamics. 
Recently we have 
studied the influence of the stochastic dynamics on the microscopic 
structure and mobility of Ising and solid-on-solid (SOS) interfaces 
that move under two types of Glauber dynamics \cite{RIKV00,RIKV01,RIKV02}. 
Both Ising and SOS interfaces are described by the Ising Hamiltonian 
\begin{equation}
{\cal H} = -\sum_{x,y} s_{x,y} \left( J_x s_{x+1,y} + J_y s_{x,y+1} 
+ H \right) 
\;, 
\label{eq:ham}
\end{equation}
where $s_{x,y} = \pm 1$ is an Ising spin at lattice site $(x,y)$,
$\sum_{x,y}$ runs over all sites on a square lattice, 
and $J_x$ and $J_y$ are ferromagnetic 
interactions in the $x$- and $y$-directions, respectively. The quantity 
$H$ is the applied ``field,'' and the interface is introduced by fixing 
$s_{x,y}=+1$ and $-$1 for large negative and positive $y$, respectively.  
We take $H \ge 0$, such 
that the interface on average moves in the positive $y$ direction. 
(This model is equivalent to a lattice-gas model with local
occupation variables $c_{x,y} \in \{0,1\}$, see details in Ref.~\cite{RIKV02}.)
The difference between the two interface types is that the Ising interface 
allows overhangs and bubbles, while these are forbidden in the SOS interface. 
However, at low and intermediate temperatures overhangs and bubbles in an 
Ising interface are rare, and a short interface segment is likely to be 
indistinguishable from an SOS interface. 
A typical SOS interface is illustrated in Fig.~\ref{fig:pict}. 

In addition to bubbles that are generated {\it at\/} the interface by
pinching off of protrusions or indentations, Ising models
in an applied field can also contain bubbles created by homogeneous
nucleation in the bulk phases \cite{DEVI92,RAMO99,NOVO00}. While such
bubbles destroy the integrity of the interface at very strong fields,
they have only a minor influence on the mobility at moderate fields 
\cite{RAMO99,NOVO00}. In this paper, like in Ref.~\cite{RIKV00}, we
exclude such nucleated bubbles by setting the transition rate
equal to zero for sites that have no neighbors with the opposite spin 
direction. As a consequence, the bulk phases
far from the interface are uniform. 

The interface dynamic is defined by the set of single-spin transition 
probabilities, $W(s_{x,y} \rightarrow -s_{x,y})$, 
and time is measured in units of attempted MC updates per site (MC steps 
per site, or MCSS). 
The first one of the dynamics used in the aforementioned 
studies \cite{RIKV00,RIKV01} is the standard discrete-time Glauber dynamic 
with the transition probability \cite{LAND00} 
\begin{equation}
W_{\rm G}(s_{x,y} \rightarrow -s_{x,y}) 
= \left[ 1 + e^{ \beta \Delta E } \right]^{-1} ,
\label{eq:glau}
\end{equation}
where $\Delta E$ is the total energy change that would result from the 
transition. 
Although $\Delta E$ can be written as a sum of the energy change 
$\Delta E_J$, due to the change in the interaction part of the Hamiltonian, 
and $\Delta E_H$, due to the change in the field energy, 
this transition rate itself cannot be factorized into a product of parts that 
depend only on $\Delta E_J$ and $\Delta E_H$, respectively. This dynamic is 
therefore classified as ``hard'' in the literature on driven particle systems 
\cite{MARR99}. 

The second type of dynamics is defined by transition 
probabilities that factorize into a part that depends only on $\Delta E_J$ 
and one that depends only on $\Delta E_H$. Such dynamics are known as ``soft'' 
\cite{MARR99}. In our recent study of a driven SOS interface 
with soft dynamics we used, for reasons of mathematical 
convenience, the ``soft Glauber dynamic'' \cite{RIKV01} 
in which each of the two parts has the Glauber form: 
\begin{equation}
W_{\rm SG}(s_{x,y} \rightarrow -s_{x,y}) 
= 
\left[ 1 + e^{\beta \Delta E_H } \right]^{-1} 
\cdot
\left[ 1 + e^{\beta \Delta E_J } \right]^{-1} 
. 
\label{eq:SG}
\end{equation}
Soft dynamics (usually with the field part proportional to a Metropolis 
transition rate \cite{GUO90,KOTR91}) 
are often used for lattice-gas simulations, 
in which the field term corresponds to the 
entropic part of a chemical-potential difference. 
In Ref.~\cite{RIKV01} we showed, in agreement with a theoretical prediction 
in Ref.~\cite{RIKV00}, that the soft dynamic leads to a microscopic 
SOS interface 
structure that is identical to the equilibrium interface in zero field, 
irrespective of the value of the applied field. This is in contrast with 
the result for hard dynamics \cite{RIKV02}, which lead to an intrinsic 
interface width that increases dramatically with the field. The purpose of 
the present paper is to study the effects of the soft Glauber dynamic for 
an Ising interface (which may contain bubbles and overhangs) and compare those 
with the cases of an Ising interface with the standard (hard) Glauber dynamic 
\cite{RIKV00} and SOS interfaces with the soft Glauber dynamic \cite{RIKV01} 
and hard Glauber dynamic \cite{RIKV02}. 

The rest of this paper is organized as follows. 
Theoretical results for the interface structure and velocity are
surveyed in Sec.~\ref{sec:ISV}. Comparisons with extensive dynamic MC
simulations are given in Sec.~\ref{sec:MC}, 
with results for the interface velocity in Sec.~\ref{sec:MCa} 
and for the interface structure and skewness in Sec.~\ref{sec:MCc}. 
Our conclusions are drawn in Sec.~\ref{sec:Concl}.

\section{Interface Structure and Velocity}
\label{sec:ISV}

\begin{figure}[ht] 
\includegraphics[angle=0,width=.50\textwidth]{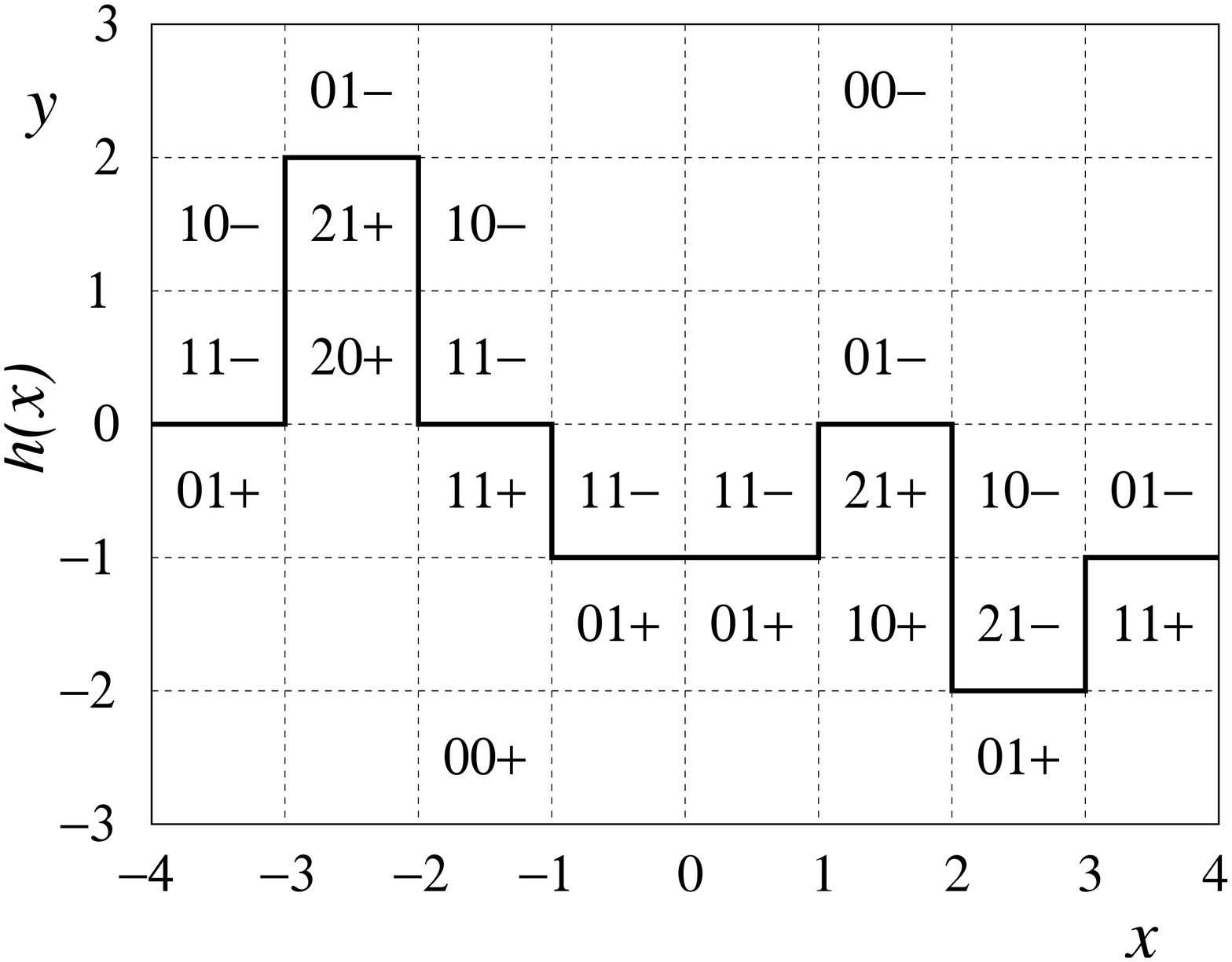} 
\caption[]{
A short segment of an SOS interface $y=h(x)$ between a positively
magnetized phase below and a negative phase
above. The step heights are $\delta(x) = h(x+1/2) - h(x-1/2)$. Interface
sites representative of the different SOS spin classes (see
Table~\protect\ref{table:class} 
and Table~\protect\ref{table:class2}) are marked with the
notation $jks$ explained in the text. Sites in the uniform bulk phases are
$00-$ and $00+$. From Ref.~\protect\cite{RIKV02}. 
}
\label{fig:pict}
\end{figure}

With the Ising Hamiltonian there is only a finite 
number of different values of $\Delta E$, and the spins can therefore be 
divided into classes \cite{BORT75,SPOH93,NOVO95A}, 
labeled by the spin value $s$ and the number 
of broken bonds between the spin and its nearest neighbors in the $x$ and 
$y$ direction, $j$ and $k$, respectively. 
The eighteen different Ising spin classes are denoted $jks$ 
with $j,k \in \{0,1,2\}$. They are listed in Table~\ref{table:class}, 
and subsets are also listed in Table~\ref{table:class2}
and shown in Fig.~\ref{fig:pict}. 

The Burton-Cabrera-Frank  
SOS model \cite{BURT51} considers an interface in a lattice gas or $S=1/2$ 
Ising system on a square lattice of unit lattice constant 
as a single-valued integer function $h(x)$ of the $x$-coordinate, with
steps $\delta(x) = h(x+1/2) - h(x-1/2)$ at integer values of $x$. 
A typical SOS interface configuration is shown in Fig.~\ref{fig:pict}. 
The heights of the individual 
steps are assumed to be statistically independent and, in the case of a flat 
interface, identically distributed. These assumptions are exact in equilibrium
\cite{BURT51}.
The step-height probability density function (PDF) 
is given by the interaction energy corresponding to the $|\delta(x)|$ broken 
$J_x$-bonds between spins in the columns centered 
at $(x-1/2)$ and  $(x+1/2)$ as  
\begin{equation}
p[\delta(x)] = Z(\phi)^{-1} X^{|\delta(x)|}
\ e^{ \gamma(\phi) \delta(x) } \;. 
\label{eq:step_pdf}
\end{equation}
The Boltzmann factor $X = e^{- 2 \beta J_x}$ determines the width of the PDF, 
and the Lagrange multiplier $\gamma(\phi)$ maintains the mean step 
height at an $x$-independent value, $\langle \delta(x) \rangle = \tan \phi$, 
where $\phi$ is the overall angle between the interface and the $x$ axis. 
The Lagrange multiplier is given by 
\begin{equation}
e^{\gamma (\phi)} 
= 
\frac{ \left(1+X^2 \right)\tan \phi 
+ \left[ \left( 1 - X^2 \right)^2 \tan^2 \phi + 4 X^2 \right]^{1/2}}
{2 X \left( 1 + \tan \phi \right)} 
\;.
\label{eq:chgam}
\end{equation}
The partition function for the step height $\delta(x)$ is 
\begin{equation}
Z(\phi)
=
\sum_{\delta = -\infty}^{+\infty} X^{|\delta|} e^{ \gamma(\phi) \delta } 
= 
\frac{1-X^2}{1 - 2 X \cosh \gamma(\phi) + X^2}
\;. 
\label{eq:Z}
\end{equation}
(See details in Refs.~\cite{RIKV00,RIKV02}). 
Simple results are obtained for 
$\phi = 0$, which yields $\gamma(0) = 0$ and 
\begin{equation}
Z(0) = (1+X)/(1-X) \;,
\label{eq:Z0}
\end{equation} 
and for $\phi = 45^\circ$, which yields 
\begin{equation}
e^{\gamma (45^\circ)} 
= 
(1+X^2)/2X
\label{eq:chgam45}
\end{equation}
and
\begin{equation}
Z(45^\circ) 
=
2 (1+X^2)/(1-X^2) \;.
\label{eq:Z45}
\end{equation}
For soft dynamics (but not for hard dynamics), $X$ remains independent of 
$H$ when the system is driven away from equilibrium \cite{RIKV00,RIKV01,RIKV02}.

The mean spin-class 
populations, $\langle n(jks) \rangle$, are all obtained from the 
product of the independent PDFs for $\delta(x)$ and $\delta(x$+1). 
Symmetry of $p[\delta(x)]$ under the transformation 
$(x,\phi,\delta) \rightarrow (-x,-\phi,-\delta)$ ensures that 
$\langle n(jk-) \rangle = \langle n(jk+) \rangle$ for all $j$ and $k$. 
Numerical results illustrating the breakdown of this up/down symmetry for 
large $|H|$ are discussed in Sec.~\ref{sec:MCc}. 
As discussed in Ref.~\cite{RIKV00}, calculation of the individual class 
populations 
is straightforward but somewhat tedious, especially for nonzero $\phi$. 
The final results are summarized in Table~\ref{table:class2}. 

Whenever a spin  flips from $-$1 to +1, 
the corresponding column of the interface advances by one lattice constant 
in the $y$ direction. Conversely, the column 
recedes by one lattice constant when a spin 
flips from +1 to $-$1. The corresponding energy changes are 
given in the third and fourth columns of Table~\ref{table:class}. 
Since the spin-class populations on both sides of the 
interface are equal in this approximation, the contribution 
to the mean velocity in the $y$ direction 
from sites in the classes $jk-$ and $jk+$ becomes 
\begin{equation}
\langle v_y(jk) \rangle 
= 
W \left( \beta \Delta E(jk-) \right)
-
W \left( \beta \Delta E(jk+) \right) 
 \;. 
\label{eq_generalv}
\end{equation}
The results corresponding to 
the soft Glauber transition probabilities used here, Eq.~(\ref{eq:SG}), 
are listed in the last column of Table~\ref{table:class2}.  
The mean propagation velocity perpendicular to the interface becomes  
\begin{equation}
\langle v_\perp (T,H,\phi) \rangle 
= 
\cos \phi \sum_{j,k} \langle n(jks) \rangle \langle v_y (jk) \rangle 
\;, 
\label{eq:totalv}
\end{equation}
where the sum runs over the classes included in Table~\ref{table:class2}. 
While the general result is cumbersome 
if written out in detail, using the fact that $e^{-4 \beta J_x} = X^2$ 
for the soft Glauber dynamic \cite{RIKV00,RIKV01,RIKV02}, 
we obtain relatively compact formulas 
for the special cases of $\phi=0$ and $\phi = 45^\circ$: 
\begin{eqnarray}
\langle v_\perp (T,H,0) \rangle 
&=&
X
\left\{
\frac{1}{1+X^2} 
+ 
\frac{X}
{(1+X)^2 (1-X^2)}
\left[
\frac{2(1+2X)}{1+e^{4 \beta J_y}}
+
\frac{X^2}{1+X^2 e^{4 \beta J_y}}
\right] 
\right\} 
{\tanh (\beta H)} \nonumber\\
&&
\label{eq:totalv0}
\end{eqnarray}
and
\begin{eqnarray}
\langle v_\perp (T,H,45^\circ) \rangle 
&=&
\left\{
\frac{1}{2} + \frac{2X^2}{(1+X^2)^2} 
+ 
\frac{1}{1-X^4}\left[\frac{1+2X^2+3X^4}{1+e^{4 \beta J_y}} 
+ 
\frac{X^4}{1+X^2e^{4 \beta J_y}} \right] 
\right\} 
\frac{\tanh (\beta H)}{2 \sqrt{2}} \;. \nonumber\\
&&
\label{eq:totalv45}
\end{eqnarray}

\section{Comparison with Monte Carlo Simulations} 
\label{sec:MC} 

We have compared the analytical estimates of the propagation velocities 
and spin-class populations developed above with MC simulations of the 
same model for $J_x = J_y = J$. The details of our implementation 
of the discrete-time $n$-fold way rejection-free MC 
algorithm \cite{BORT75} are described in Ref.~\cite{RIKV00}. 
By keeping only the interface sites in memory, the algorithm is not subject 
to size limitations in the $y$ direction, enabling simulations 
for arbitrarily long times. 

Our numerical results are based on simulations with  
$L_x = 10\,000$ and $\phi$ between 0 and $45^\circ$ 
for several temperatures below $T_c$. ($T_c = -2J/\ln(\sqrt{2} -1)
\approx 2.269J$ is the critical temperature for the isotropic,
square-lattice Ising model \cite{ONSA44}.) 
In order to ensure stationarity we ran the simulation for 10\,000 
$n$-fold way updates per updatable spin (UPS) 
before taking any measurements, and the results were averaged over 
200\,000 UPS \cite{ENDNOTE}. 

\subsection{Interface velocities}
\label{sec:MCa}

\begin{figure}[ht] 
\includegraphics[angle=0,width=.50\textwidth]{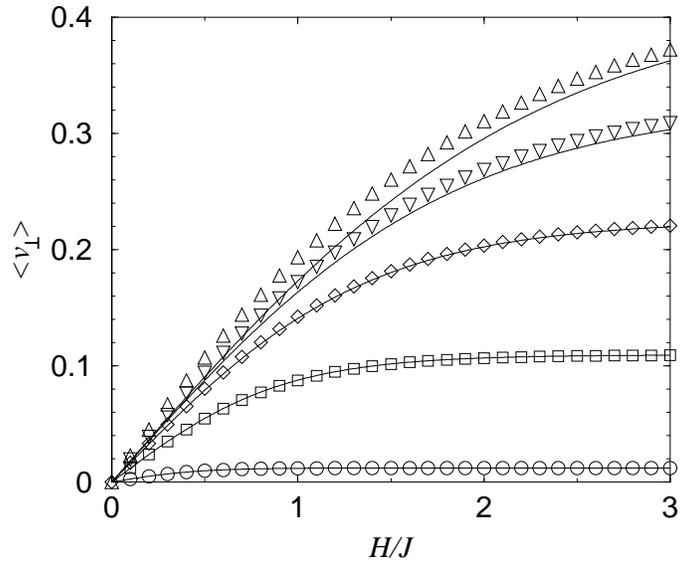} 
\caption[]{
The average stationary normal interface velocity $\langle v_\perp \rangle$, 
shown vs $H$ for $\phi = 0$. The MC results are shown as data points, and 
the theoretical predictions as solid curves. From below to above, the  
temperatures are $T/T_c = 0.2$, 0.4, 0.6, 0.8, and 1.0.
}
\label{fig:vvsH}
\end{figure}

First we compare the simulated interface velocities 
with the analytical approximation, Eq.~(\ref{eq:totalv}). 
Figure~\ref{fig:vvsH} shows the normal velocity vs $H$ for 
$\phi = 0$ and a range of temperatures up to $T_c$. 
There is excellent agreement between the MC results and the theory 
for temperatures below 0.8$T_c$. 

\begin{figure}[ht]
\includegraphics[angle=0,width=.50\textwidth]{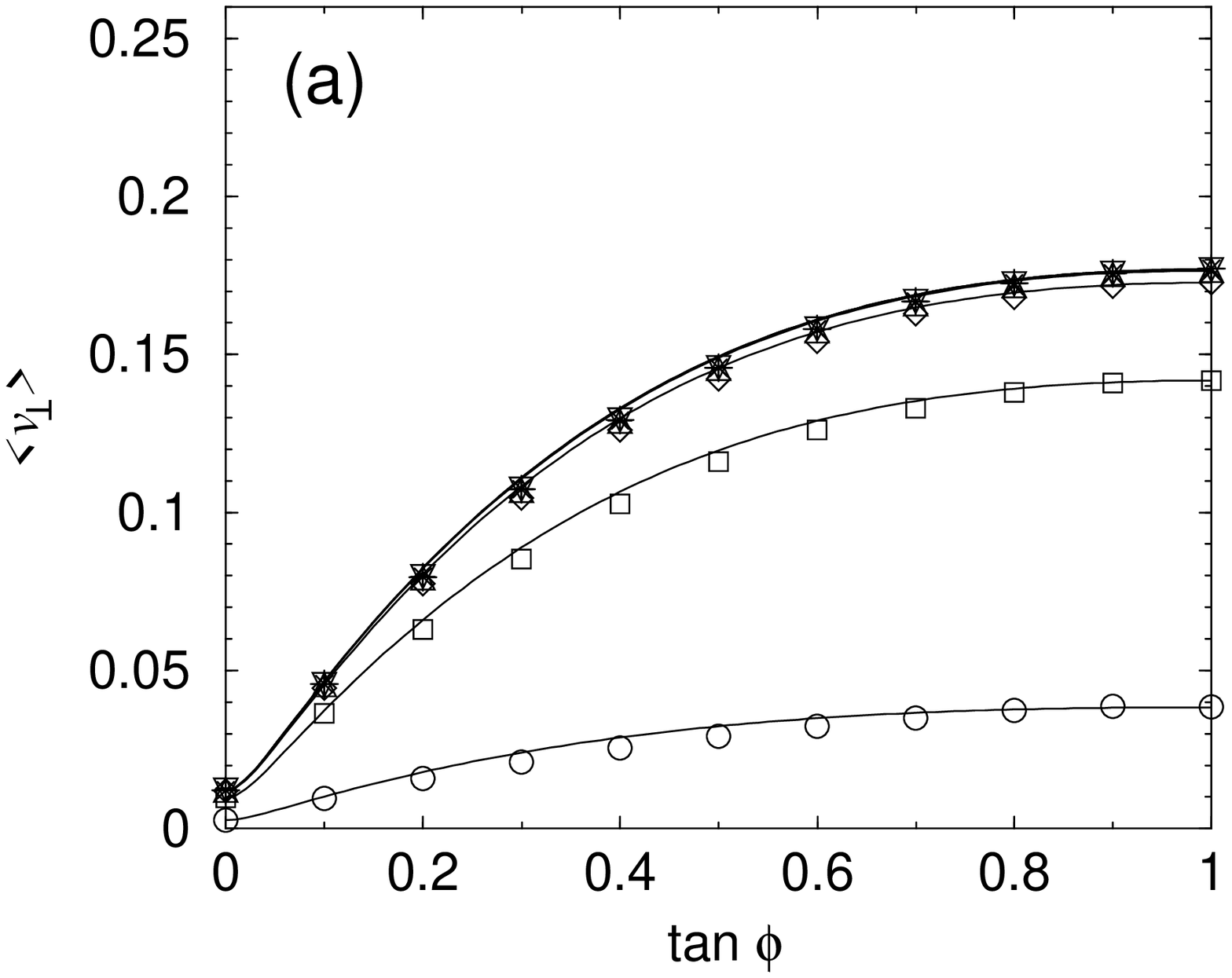} 
\includegraphics[angle=0,width=.50\textwidth]{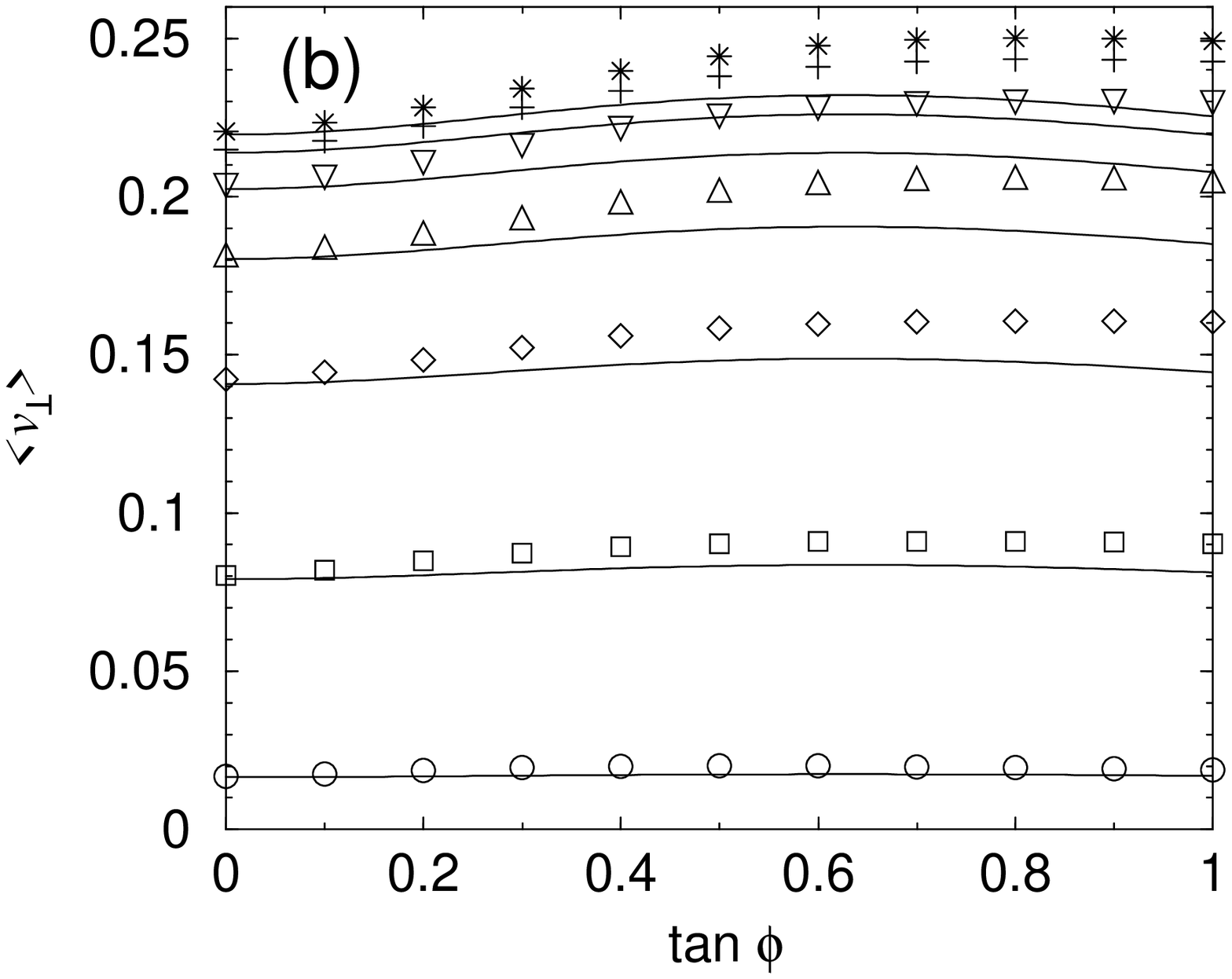} 
\caption[]{
The average stationary normal interface velocity $\langle v_\perp \rangle$, 
shown vs $\tan \phi$ for (from below to above) 
$H/J = 0.1$, 0.5, 1.0, 1.5, 2.0, 2.5, and 3.0.
The MC results are shown as data points, and 
the theoretical predictions as solid curves. 
(a) $T=0.2T_c$. 
(b) $T=0.6T_c$. 
}
\label{fig:vvsA}
\end{figure}

The dependence of the normal velocity on the tilt angle $\phi$ is shown in 
Fig.~\ref{fig:vvsA} for several values of $H/J$ between 0.1 and 3.0. 
At $T=0.2T_c$ the velocity increases with 
$\phi$ in agreement with the polynuclear growth model 
\cite{DEVI92,KRUG89,KERT89} 
at small angles and the single-step model for larger angles 
\cite{DEVI92,SPOH93,MEAK86,PLIS87} 
[Fig.~\ref{fig:vvsA}(a)]. At $T=0.6T_c$, on the other hand, 
the velocity is nearly isotropic, with a weak increase with $\phi$ for the 
strongest fields [Fig.~\ref{fig:vvsA}(b)]. 
For the lowest temperature the 
agreement between the simulations and the analytical results is excellent 
everywhere. For the higher temperature it is also reasonable, but 
better for weak than for strong fields. 

\begin{figure}[ht] 
\includegraphics[angle=0,width=.50\textwidth]{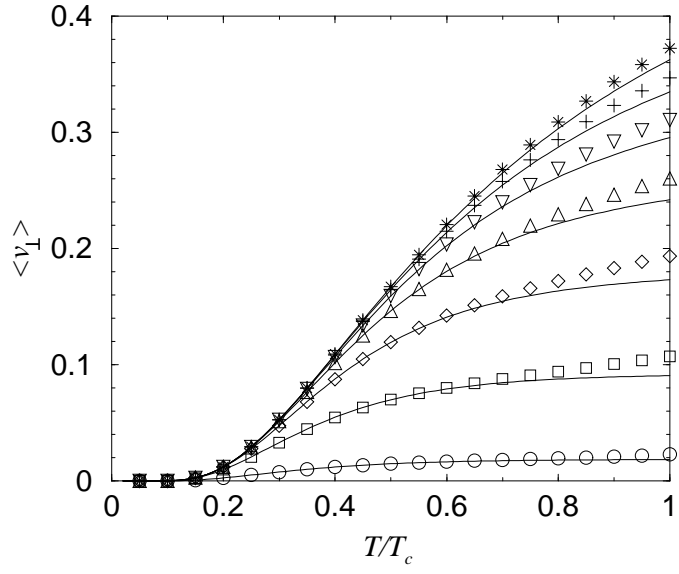} 
\caption[]{
The average stationary normal interface velocity $\langle v_\perp \rangle$, 
shown vs $T$ at $\phi = 0$ for (from below to above) 
$H/J = 0.1$, 0.5, 1.0, 1.5, 2.0, 2.5, and 3.0.
The MC results are shown as data points, and 
the theoretical predictions as solid curves. 
}
\label{fig:vvsT}
\end{figure}

The temperature dependence of the normal interface velocity is shown in 
Fig.~\ref{fig:vvsT} for several values of $H/J$ between 0.1 and 3.0. 
The agreement between the simulations and the analytical results is excellent 
except for combinations of high temperatures and strong fields. 
In contrast to the results for hard dynamics (see Fig.~5 of Ref.~\cite{RIKV00} 
for Ising interfaces and Fig.~8 of Ref.~\cite{RIKV02} for SOS interfaces), the 
velocity goes to zero at $T=0$ for {\it all\/} values of $H$, 
not just for $H/J < 2$. 
This result agrees with our finding for the SOS model with soft
dynamics \cite{RIKV01}. 
As predicted by the theoretical results in Ref.~\cite{RIKV00}, 
there is thus {\it no\/} discontinuity in the interface 
velocity at $T=0$ and $H/J = 2$ for soft dynamics. 

\subsection{Spin-class populations and skewness}
\label{sec:MCc}

A closer look at the performance of the mean-field approximation for the 
interface structure  
is provided by the mean spin-class populations. The analytical predictions 
for the class populations are 
based on the assumption that different steps are statistically independent. 
A comparison of the simulation results with the analytical 
predictions therefore gives a way of testing this assumption. 

\begin{figure}[ht] 
\includegraphics[width=.48\textwidth]{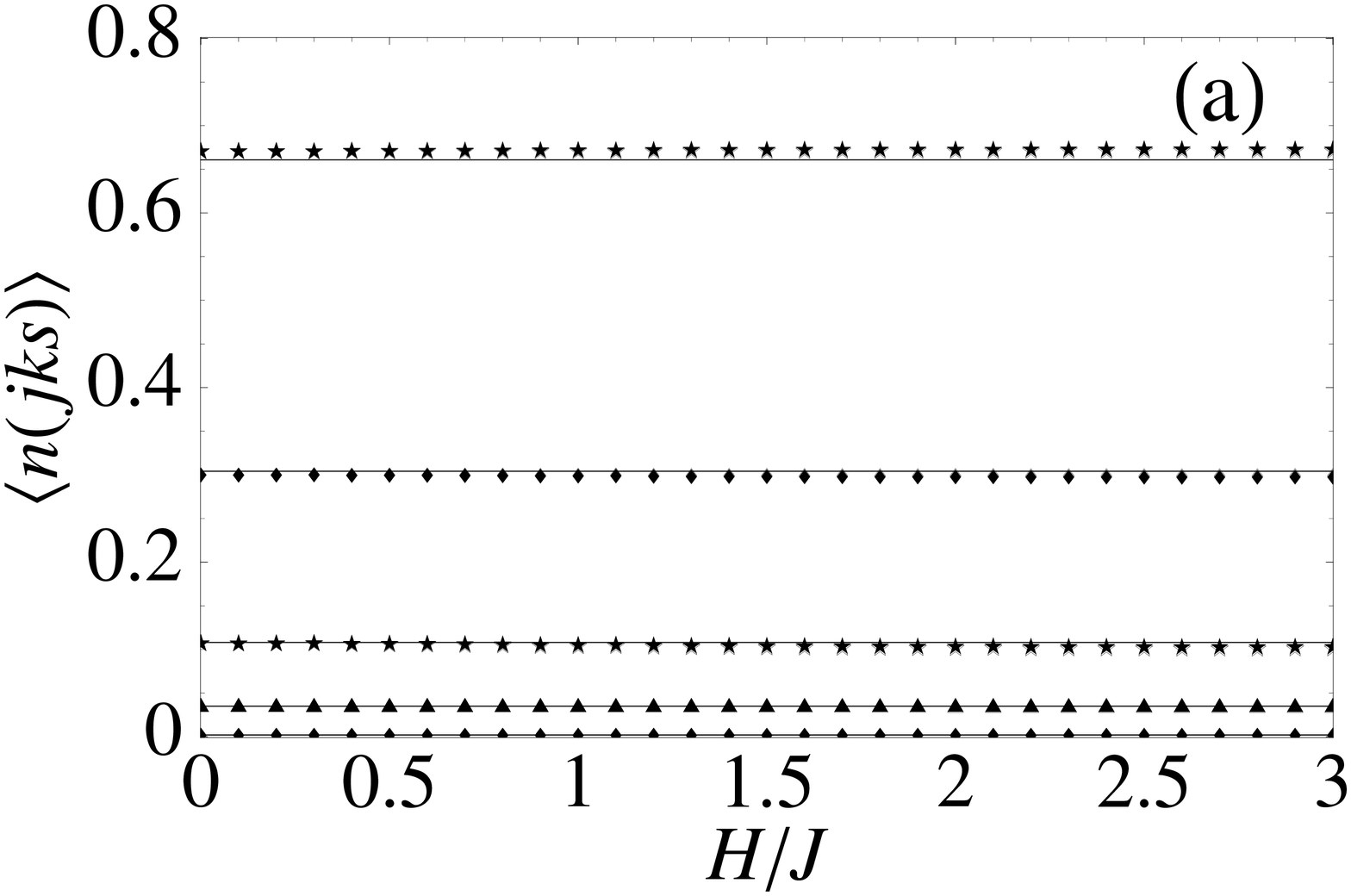} 
\includegraphics[width=.48\textwidth]{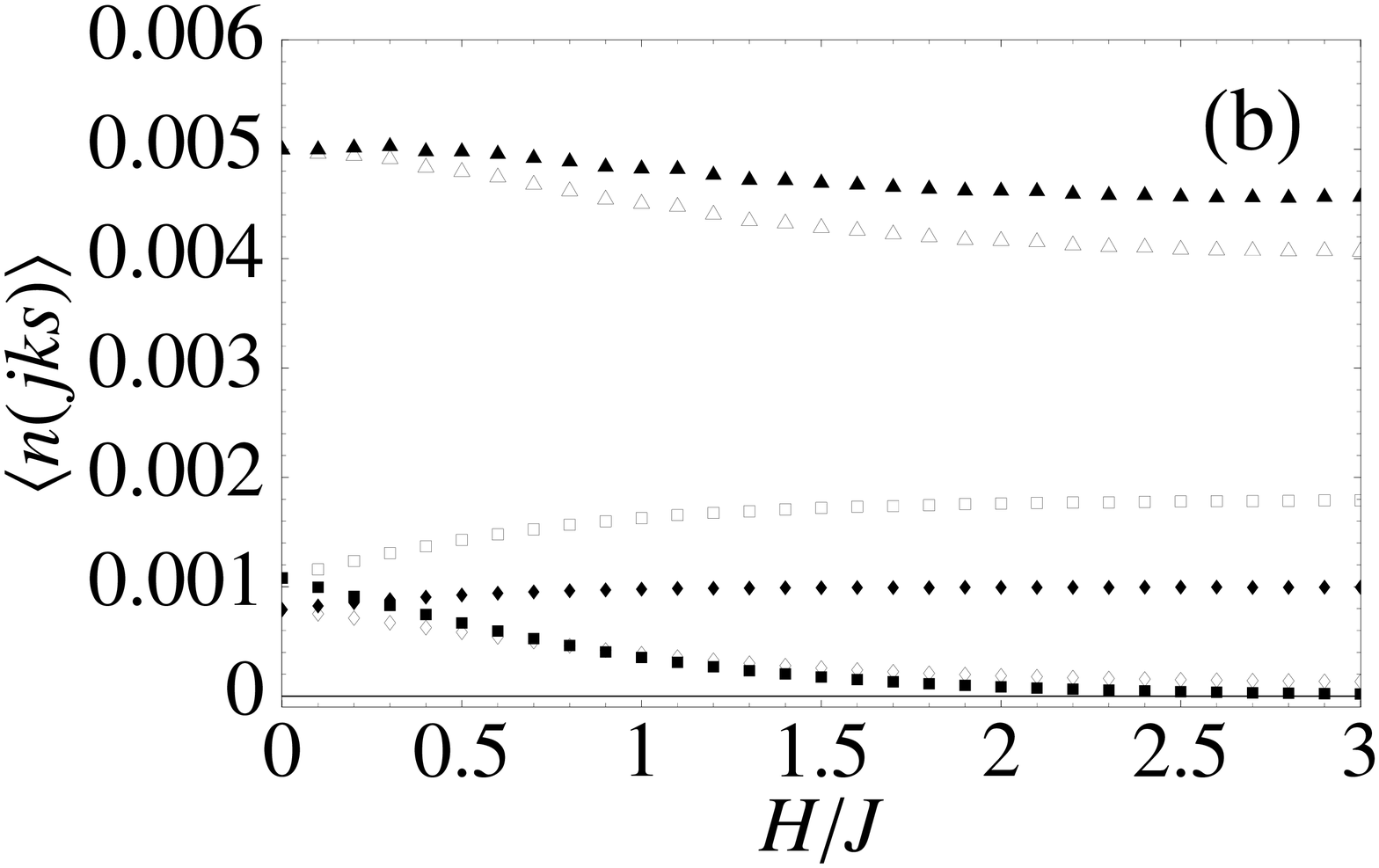} 
\includegraphics[width=.48\textwidth]{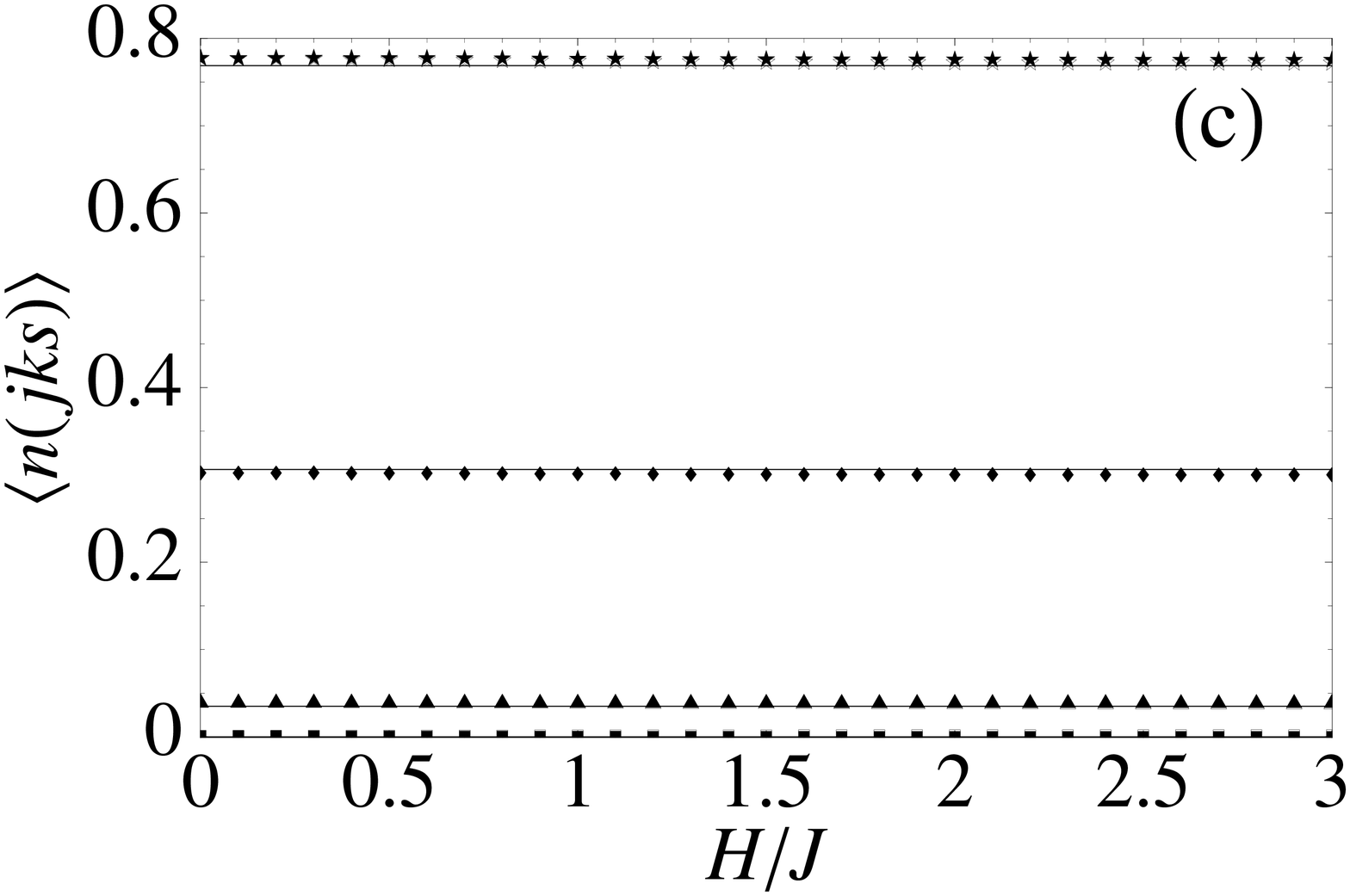} 
\caption[]{
Mean stationary class populations $\langle n(jks) \rangle$, shown vs $H$ at 
$\phi = 0$ and $T = 0.6T_c$. 
Solid lines correspond to the theoretical predictions, while 
filled symbols denote $s=+1$ and empty symbols (in most cases hidden by 
the corresponding filled symbols) denote $s=-1$. 
(a)
The ten SOS-compatible classes, from top to bottom 
01$s$, 11$s$, 10$s$, 21$s$, and 20$s$. 
(b) 
The six classes with two broken $y$-bonds, which have zero populations in 
the SOS approximation, 12$s$ (triangles), 22$s$ (squares), and 
02$s$ (diamonds). 
(c)
From above to below are shown the aggregate populations of classes with 
one (01$s$ and 10$s$), two (11$s$, 20$s$, and 02$s$), 
three (21$s$ and 12$s$) and four (22$s$) broken bonds. 
}
\label{fig:class}
\end{figure}

The ten mean class populations that have nonzero populations in the 
SOS approximation, $\langle n(01s) \rangle$, 
$\langle n(11s) \rangle$, $\langle n(10s) \rangle$, 
and $\langle n(21s) \rangle$, and $\langle n(20s) \rangle$ 
with $s = \pm 1$ are shown vs $H$ 
in Fig.~\ref{fig:class}(a) for $\phi = 0$ and $T=0.6T_c$. 
Filled symbols represent
$s=+1$, while empty symbols (almost completely hidden behind the corresponding 
filled symbols) represent $s=-1$. 
The class populations are practically independent of $H$, in 
agreement with the theoretical prediction for the soft dynamic 
\cite{RIKV00,RIKV01}, and in contrast to the result for the Ising model
with hard Glauber dynamics (see Fig.~7(a) of Ref.~\cite{RIKV00}). 
Deviations from the SOS approximation are indicated by the 
nonzero populations in the classes with two broken $y$-bonds, 
$\langle n(12s) \rangle$, $\langle n(22s) \rangle$, 
and $\langle n(02s) \rangle$, which are 
shown in Fig.~\ref{fig:class}(b). These populations are only of the order of 
10$^{-3}$, about two orders of magnitude less than for the hard Glauber 
dynamic (see Fig.~7(b) of Ref.~\cite{RIKV00}), and they 
show significant differences between the two spin values (see below). 
Figure~\ref{fig:class}(c) shows the combined populations in classes with 
one, two, three, and four broken bonds, respectively. The results are 
dominated by the SOS-compatible classes and show good agreement between 
simulations and theory. 

\begin{figure}[ht] 
\includegraphics[angle=0,width=.50\textwidth]{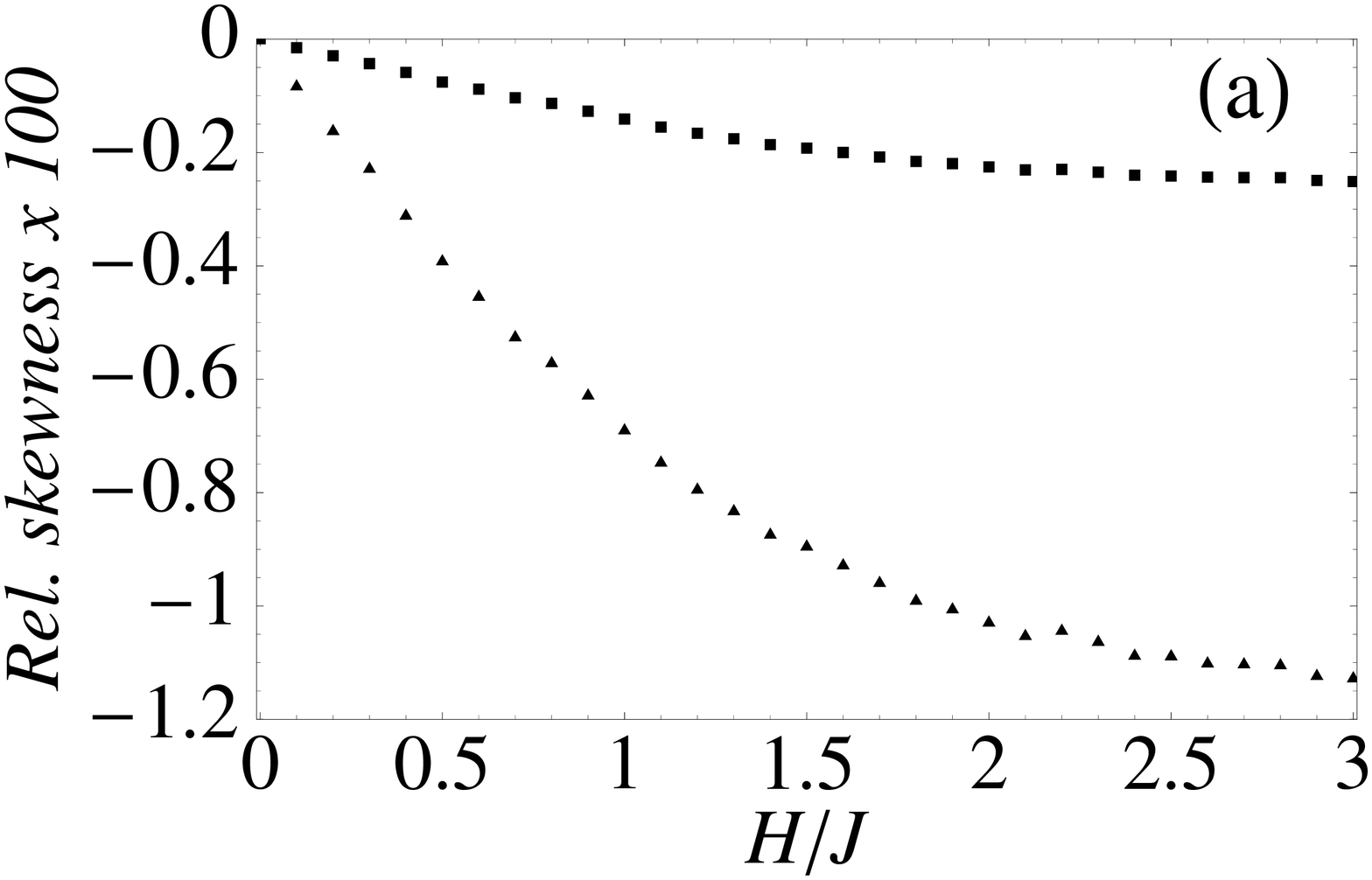} 
\includegraphics[angle=0,width=.50\textwidth]{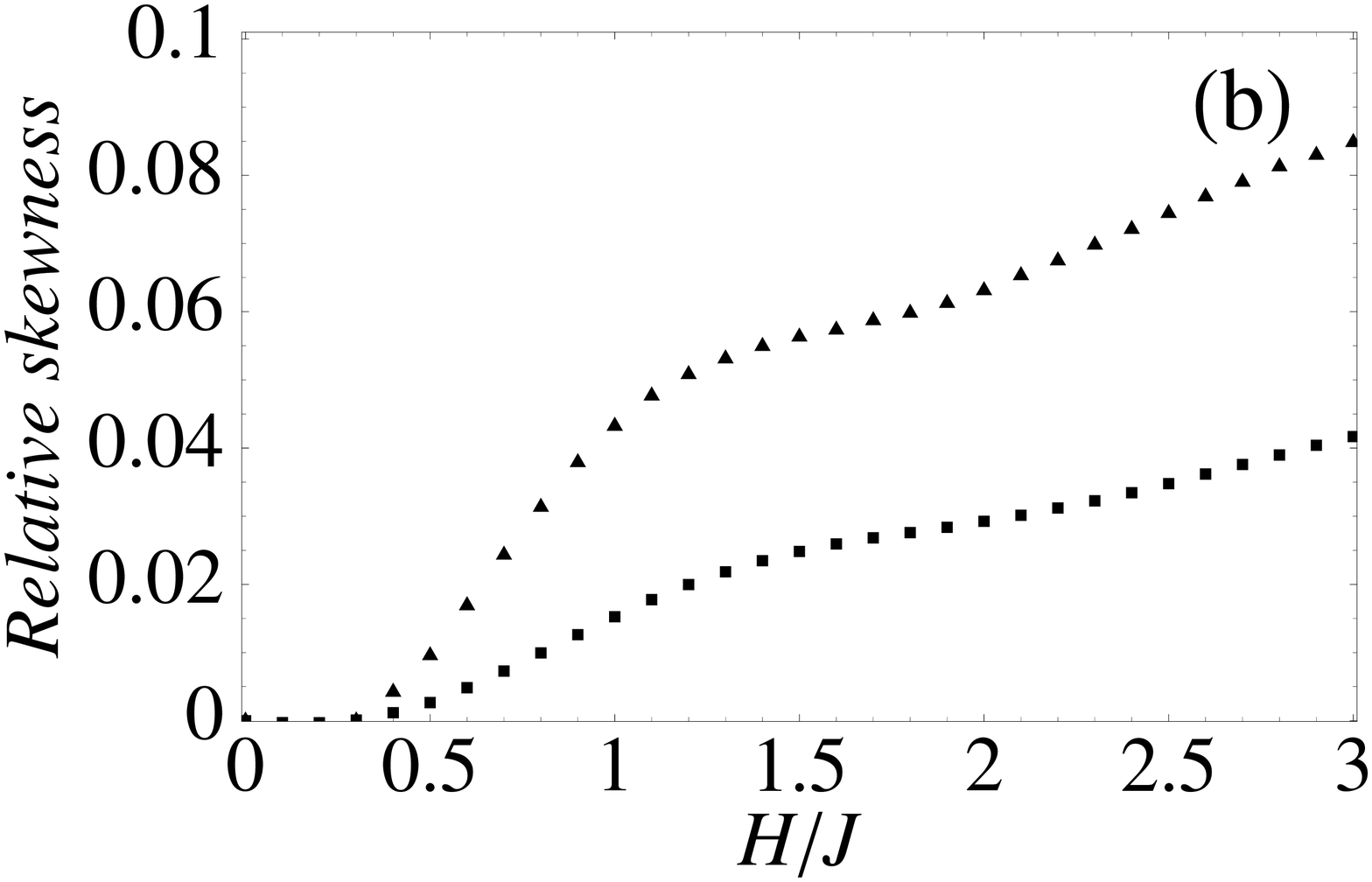} 
\includegraphics[angle=0,width=.50\textwidth]{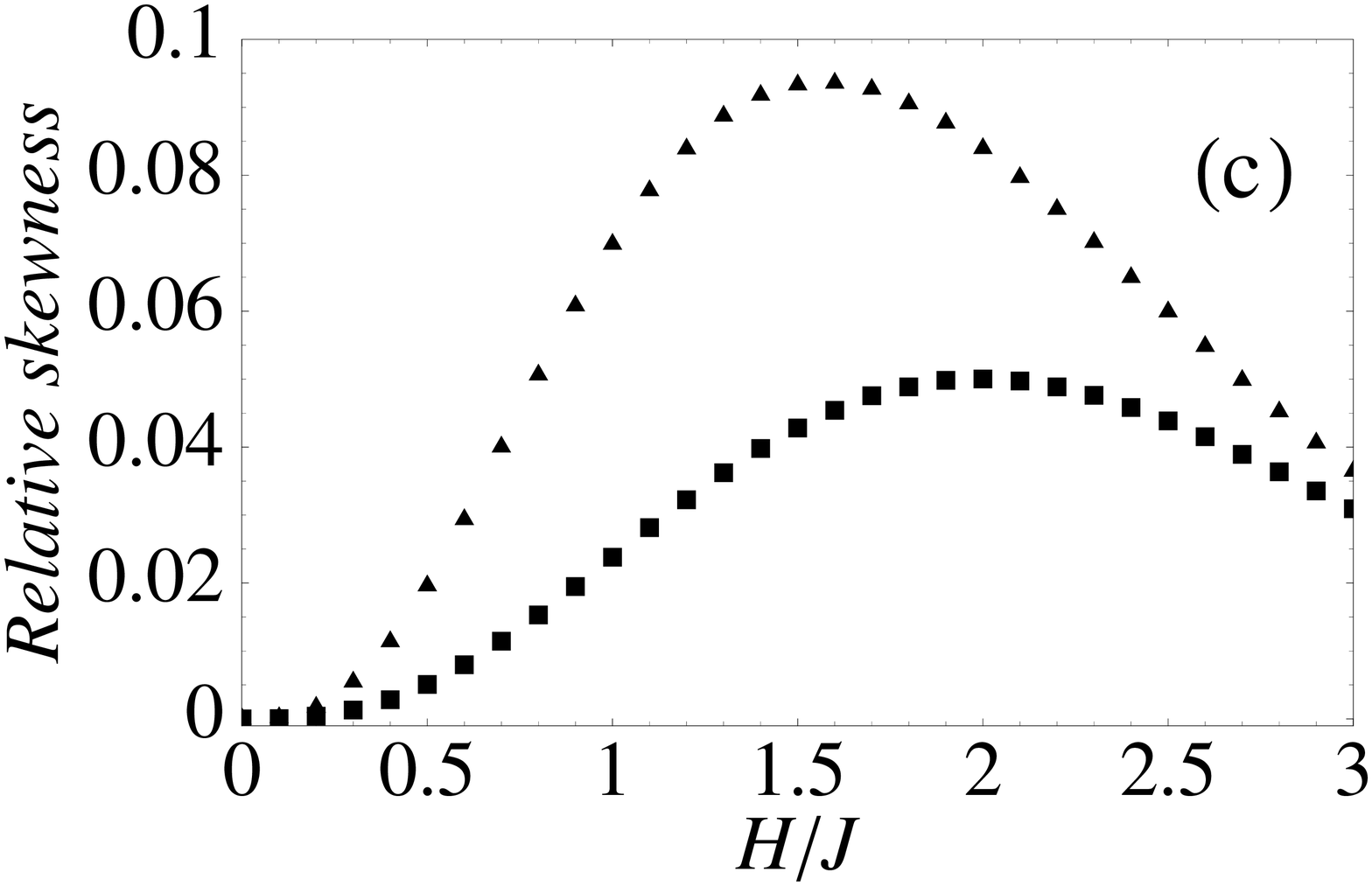} 
\caption[]{
The two relative skewness parameters $\rho$ (triangles) and 
$\epsilon$ (squares), defined in Eqs.~(\ref{eq:rho})
and~(\ref{eq:epsi}), respectively, 
shown vs $H$ for $\phi = 0$ at $T=0.6T_c$.
(a)
The skewness parameters multiplied by 100 for the Ising model 
with soft Glauber dynamics, discussed in this paper. 
(b)
The skewness parameters (no multiplication) 
for the Ising model with hard Glauber dynamics, 
discussed in Ref.~\cite{RIKV00}. 
(c)
The skewness parameters (no multiplication) 
for the SOS model with hard Glauber dynamics,
from Ref.~\cite{RIKV02}. 
Note the different sign and the scale difference of two 
orders of magnitude between the soft dynamic shown in (a) 
and the hard dynamics shown in (b) and (c). 
}
\label{fig:skew}
\end{figure}

The skewness between the spin populations on the leading and trailing edges 
of the interface are a consequence of short-range correlations between 
neighboring steps, and it is quite commonly observed in driven interfaces. 
This is the case, 
even when the {\it long-range\/} correlations vanish as they do
for interfaces in the 
KPZ dynamic universality class \cite{BARA95,KARD86}, 
to which the present model belongs. Skewness has also been 
observed in several other SOS-type models \cite{NEER97,PIER99,KORN00}. 
The correlations associated with the skewness generally lead to a broadening 
of protrusions on the leading edge (``hilltops''), while 
those on the trailing edge (``valley bottoms'') are sharpened \cite{NEER97}, 
or the other way around \cite {KORN00}. In terms of spin-class populations, 
the former corresponds to $\langle n(21-) \rangle > \langle n(21+) \rangle$ 
and $\langle n(11+) \rangle > \langle n(11-) \rangle$. The relative skewness 
can therefore be quantified by the two functions, 
\begin{equation}
\rho = \frac{\langle n(21-) \rangle - \langle n(21+) \rangle}
{\langle n(21-) \rangle + \langle n(21+) \rangle}
\;,
\label{eq:rho}
\end{equation}
introduced by Neergaard and den~Nijs \cite{NEER97}, and 
\begin{equation}
\epsilon = \frac{\langle n(11+) \rangle - \langle n(11-) \rangle}
{\langle n(11+) \rangle + \langle n(11-) \rangle}
\;.
\label{eq:epsi}
\end{equation}
These two skewness parameters for the current system 
are shown together versus $H$ in Fig.~\ref{fig:skew}(a). 
The skewness is very weak, but it is not zero 
in contrast to the SOS model with soft Glauber dynamic \cite{RIKV01}. 
Both these skewness parameters have the opposite 
sign and are about two orders of magnitude smaller than in 
the Ising model with the hard Glauber dynamic \cite{RIKV00}
[Fig.~\ref{fig:skew}(b)] and in 
the SOS model with the same hard dynamic \cite{RIKV02} 
[Fig.~\ref{fig:skew}(c)].  

\begin{figure}[ht] 
\includegraphics[angle=0,width=.50\textwidth]{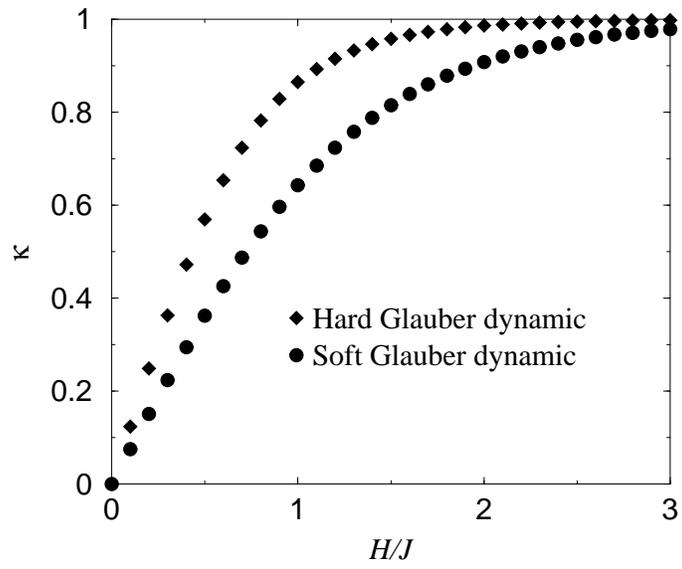} 
\caption[]{
The relative skewness parameter $\kappa$,  
shown vs $H$ for $\phi = 0$ at $T=0.6T_c$.
It measures the asymmetry
between the populations of bubbles behind and in front of the Ising interface. 
}
\label{fig:kap}
\end{figure}

The skewness parameters $\rho$ and $\epsilon$ depend on spin classes
which have nonzero populations in the SOS picture, and they can
therefore be applied to both Ising and SOS interfaces. In the Ising 
case, however, much more pronounced differences are seen in in the
populations of those classes that are not populated in the SOS model
[see Fig.~\ref{fig:class}(b)]. For Ising models, these classes can be
used to define further skewness parameters, such as 
\begin{equation}
\kappa = \frac{\langle n(22-) \rangle - \langle n(22+) \rangle}
{\langle n(22-) \rangle + \langle n(22+) \rangle}
\;.
\label{eq:kap}
\end{equation}
Here class 22$-$ represents isolated bubbles of the metastable phase
that persist as a ``wake'' behind the moving interface, while 22$+$
corresponds to a ``bow wave'' 
of bubbles of the stable phase in front of the interface,
which are created by pinching-off of protrusions. Although the
total density of such bubbles is about two orders of magnitude smaller
with the soft dynamic studied here than with hard dynamics 
(compare Fig.~\ref{fig:class}(b) 
with Fig.~7(b) of Ref.~\cite{RIKV00}), the relative asymmetry parameters 
$\kappa$ are comparable, as shown in Fig.~\ref{fig:kap}. Not
surprisingly, for strong fields the entire bubble population is found in
the wake, yielding $\kappa \alt 1$ in this limit.

\section{Conclusion}
\label{sec:Concl}

In this paper we have continued our study of the dependence of the local
structure of driven interfaces on the applied field and temperature 
and on the form of the stochastic dynamics under which they move
\cite{RIKV00,RIKV01,RIKV02}. The local interface structure is of
interest because it is it, rather than the large-scale scaling behavior,
which determines such important interface properties as mobility and chemical
reactivity. In particular, we studied the differences between interfaces
moving under {\it soft\/} stochastic dynamics, in which the influences
of changes in the field energy and the interaction energy factorize 
in the transition probabilities, and
{\it hard\/} dynamics, which do not possess such a factorization
property. 

We find that the results for the Ising
model with the soft Glauber dynamic, which is the main topic of this study,
differ relatively little from the SOS model with soft dynamics studied in
Ref.~\cite{RIKV01}. In particular, the local interface width does {\it
not\/} diverge with increasing $H$, as it does for both the Ising
\cite{RIKV00} and SOS \cite{RIKV02} models with hard dynamics. As a
result, the soft dynamics do not produce the discontinuity in the
interface velocity at $H/J = 2$ and $T=0$ that is seen for hard dynamics.   
The main qualitative difference between the Ising and SOS models with
soft dynamics is that the interface skewness for the Ising model in
nonzero field is not exactly zero, as it is for the SOS model. However,
the skewness parameters $\rho$ and $\epsilon$, which are based on
SOS-compatible spin classes, are 
about two orders of magnitude smaller  
and have the opposite sign than what is seen for the hard dynamics. 
In contrast, the relative asymmetry in the populations of bubbles behind and in
front of the interface (which do not occur in SOS models) can be  
expressed by the skewness parameter $\kappa$ and  
is comparable for the two dynamics. However, the {\it absolute\/} bubble
density is about two orders of magnitude smaller with the soft than with
the hard dynamics. 
Although a successful mean-field theory for the interface mobility of
different models and under different stochastic dynamics was developed
in Refs.~\cite{RIKV00,RIKV02}, a comparable theory that predicts the
skewness is still not available. 

Two important conclusions can be drawn from our studies. First, the
strong differences between hard and soft dynamics make it evident that
great care must be used in formulating and interpreting stochastic
models of dynamic systems. Second, experimental observation of the field
and temperature dependences of interface mobility and local interface
structure could contribute significantly to devising correct  
stochastic models of nonequilibrium physical phenomena.

\section*{Acknowledgments}
\label{sec:ACK}

P.~A.~R.\ appreciates the hospitality of the 
Department of Physics, Virginia Polytechnic Institute and State University. 
The research was supported in part 
by National Science Foundation Grant Nos.~DMR-9981815, DMR-0120310, 
and DMR-0240078, and by Florida State 
University through the Center for Materials Research and Technology and 
the School of Computational Science and Information Technology.


\clearpage

%
%
\begin{table}[ht]
\caption[]{
The spin classes in the anisotropic square-lattice Ising model. 
First column: the class labels, $jks$. 
Second column: the total energy per spin, $E(jks)$, relative 
to the state with all spins parallel and $H=0$, 
$E_0 = -2(J_x + J_y)$. 
Third column: the change in the field energy 
resulting from spin reversal from $s$ to $-s$, $\Delta E_H(jks)$.
Fourth column: the corresponding change in the 
interaction energy, $\Delta E_J(jks)$. 
In columns two and three, 
the upper (lower) sign corresponds to $s=-1$ ($s=+1$). 
The first three classes have nonzero populations in the SOS approximation, 
and flipping a spin in any of them preserves the SOS configuration. 
The two classes marked  $\dag$ also have nonzero populations in the 
SOS approximation, but flipping a spin in any of them may produce an overhang 
or a bubble. 
The two classes marked  $\ddag$ are not populated in the SOS 
approximation, but flipping a spin in any of them may produce an SOS 
configuration. 
The two classes marked  $\ast$ correspond to a bulk spin that is 
either parallel or antiparallel to all its neighbors. 
Flipping a spin in class 22$s$ yields a spin in class 00$-s$. 
The transition probabilities for all classes except 00$s$ (from which 
transitions are forbidden with the dynamic used here) are given 
by Eq.~(\ref{eq:SG}). 
}
\begin{tabular}{| l | l | l | l |}
\hline
Class, $jks$ 
& $E(jks) - E_0$ 
& $\Delta E_H(jks)$ 
& $\Delta E_J(jks)$ 
\\ 
 \hline\hline
 $01s$ 
 & $\pm H + 2J_y$ 
 & $\mp 2H$
 & $+ 4J_x $
\\
 \hline
 $11s$  
 & $\pm H + 2(J_x+J_y)$ 
 & $\mp 2H $
 & $0$
\\
 \hline
 $21s$  
 & $\pm H + 2(2J_x+J_y)$ 
 & $\mp 2H$
 & $- 4J_x $
\\
 \hline\hline
$10s$ $\dag$ 
 & $\pm H + 2J_x$ 
 & $\mp 2H$
 & $+ 4J_y $
\\
 \hline
$20s$  $\dag$
 & $\pm H + 4J_x$ 
 & $\mp 2H$
 & $- 4(J_x-J_y) $
\\
 \hline\hline
$12s$ $\ddag$ 
 & $\pm H + 2(J_x + 2J_y)$ 
 & $\mp 2H$
 & $- 4J_y $
\\
 \hline
$02s$  $\ddag$
 & $\pm H + 4J_y$ 
 & $\mp 2H$
 & $+ 4(J_x-J_y) $
\\
 \hline\hline
$22s$ $\ast$ 
 & $\pm H + 4(J_x + J_y)$ 
 & $\mp 2H$
 & $- 4(J_x + J_y) $
\\
 \hline
$00s$  $\ast$
 & $\pm H$ 
 & $\mp 2H$
 & $+ 4(J_x+J_y) $
\\
\hline
 \end{tabular}
\label{table:class}
\end{table}

\begin{table}[ht]
\caption[]{
The mean populations for the spin classes that have nonzero 
populations in the SOS approximation, 
with the corresponding contributions to the interface velocity 
under the soft 
Glauber dynamic. First column: the class labels, $jks$. 
Second column: the mean spin-class populations for 
general tilt angle $\phi$, with $\cosh \gamma(\phi)$ from Eq.~(\ref{eq:chgam}). 
Third and fourth columns: the spin-class populations for $\phi = 0$
[using $\gamma(0)=0$] and $\phi = 45^\circ$ [using Eq.~(\ref{eq:chgam45}) 
for $\exp[\gamma(45^\circ)]$], respectively. 
Fifth column: the contributions to the mean interface 
velocity in the $y$ direction from spins in classes $jk-$ and $jk+$, 
Eq.~(\protect\ref{eq_generalv}), using the soft Glauber dynamic, 
Eq.~(\ref{eq:SG}). 
}
\begin{tabular}{| l | l | l | l | l |}
\hline
Class, $jks$ 
& $\langle n(jks) \rangle$, general $\phi$ 
& $\langle n(jks) \rangle$, $\phi=0$ 
& $\langle n(jks) \rangle$, $\phi=45^\circ$ 
& $\langle v_y(jk) \rangle$ 
\\ 
 \hline\hline
 $01s$  
 & $\frac{1 - 2X \cosh \gamma (\phi) + X^2}{(1-X^2)^2}$ 
 & $\frac{1}{(1+X)^2}$ 
 & $\frac{1}{2(1+X^2)}$ 
 & $\frac{\tanh\left( \beta H \right)}{1 + e^{4 \beta J_x}}$  
\\
 \hline
 $11s$  
 & $\frac{2X[(1+X^2) \cosh \gamma (\phi) - 2X]}{(1-X^2)^2}$ 
 & $\frac{2X}{(1+X)^2}$ 
 & $\frac{1}{2}$ 
 & $\frac{\tanh\left( \beta H \right)}{2}$
\\
 \hline
 $21s$  
 & $\frac{X^2[1-2X\cosh\gamma(\phi)+X^2]}{(1-X^2)^2}$ 
 & $\frac{X^2}{(1+X)^2}$ 
 & $\frac{X^2}{2(1+X^2)}$ 
 & $\frac{\tanh\left( \beta H \right)}{1 + e^{-4 \beta J_x}}$
\\
 \hline\hline
$10s$ $\dag$ 
 & $\frac{2X^2}{1-X^2} 
\left\{
\frac{2\cosh^2\gamma(\phi)-1-2X\cosh\gamma(\phi)+X^2}
{1-2X\cosh\gamma(\phi)+X^2} \right.
$ 
 & $\frac{2X^2(1+2X)}{(1-X^2)(1+X)^2}$ 
 & $\frac{1+2X^2+3X^4}{2(1-X^4)}$ 
 & $\frac{\tanh\left( \beta H \right)}{1 + e^{4 \beta J_y}}$
\\
 &$ \left.
-
\frac{X^2 [1-2X\cosh\gamma(\phi)+X^2]}{(1-X^2)^2}
\right\}$ 
 &  
 &  
 & 
\\
 \hline
$20s$ $\dag$ 
 & $\frac{X^4 [1-2X\cosh\gamma(\phi)+X^2]}{(1-X^2)^3}$ 
 & $\frac{X^4}{(1-X^2)(1+X)^2}$ 
 & $\frac{X^4}{2(1-X^4)}$ 
 & $\frac{\tanh\left( \beta H \right)}{1 + e^{-4 \beta (J_x - J_y)}}$
\\
\hline
 \end{tabular}
\label{table:class2}
\end{table}
\clearpage 

\end{document}